# The Cosmological Foundation of Our World, seen in a Revised History of our Universe.

Tom Gehrels

Department of Planetary Sciences, University of Arizona, Tucson, AZ, USA

June 3, 2009

**Abstract**  This paper has two parts, for a specific multiverse, and for the origin of our universe as it resulted from that multiverse. The first is based on the Planck domain and a Chandrasekhar equation that have quantum, relativity, gravity, and atomic physics in unified operation. The multiverse is an evolutionary system whereby universes survive only when they have those physics, and near-critical mass such that they do not collapse, nor expand too fast.

The second part is based on 15 sets of observations of nucleosynthesis and particle properties, aging and demise for our universe, as well as of its early stages. The multiverse is supplied by debris from the aging universes, arriving on the accelerated expansion of intergalactic space. New universes accrete from the debris, which is re-energized and re-constituted gravitationally. In the process, the basic particle properties appear to have been preserved such that our universe originated $\sim 10^{37}$ Planck times later, $\sim 10^{-6}$ secs later, than it would have done in a Big Bang. Schwarzschild's limit provides a confirmation, and information on dark energy.

CONTENTS

## CONTEXT

### 1.1. INTRODUCTION

This paper has two derivations, the first is an exploration of the cosmos by using equations developed by Max Planck (1858-1947) and Subrahmanyan Chandrasekhar (1910-1995, who is referred to as "Chandra" as there already is a spacecraft thus named in his honor). The first derivation is based entirely on the equations and surprising discoveries emerge. The mass of our primordial universe is finite, and there is a multiverse with known mass, physics, and evolution of its universes.

Planck explored the physics that applies in the cosmos and this resulted in his definition of length, mass, time, and temperature in terms of the four primary cosmological constants h, c, G, and k (Planck 1899). Planck's concepts had a strong influence on Einstein as of 1914 when he was employed by Planck at Berlin University. During the great period of learning physics in the 1920s an understanding resulted of how the h, c, G, and k constants represent quantum, relativity, gravity, and atomic physics; this was accomplished through the participation of several well-known physicists.

At the very end of the 1920s, young Chandrasekhar began to develop his theory of stellar structure. In that process, he discovered an expression for cosmic masses in terms of h, c, and G, while he added H for the proton mass (Chandrasekhar 1951). That reference was also for our class notes at the secluded Yerkes Observatory of the University of Chicago where I was one of the students in the 1950s (Gehrels 2007a). Chandra published the cosmic-mass definition three times (Chandrasekhar 1937, 1951, 1989). The present paper begins where Chandra left off by comparing his expression with the Planck domain.

The second derivation answers the question "how did our universe come from the multiverse?" This discussion is simpler and shorter, but separated from the first because this is no longer based on the equations, but on observations made of aging and demise for our universe and of its early stages.

Earlier versions of this work show its development and a few blunders (Gehrels 2007b, c, d). The outline of the present paper is in the above Contents. It has two subdivisions: the first with an inter-disciplinary overview of context, and the second starts with the technical discussions in Sec. 2.1.

### 1.2. CONTEXT

We should first try to understand the context of this paper, its setting and environment. A topic is understood best when its setting is clearly defined and for this paper that is the ensemble of all universes, the multiverse. There does not seem to be any other modeling of the multiverse with well-defined numbers, masses, and physics for its universes, but there are some contexts needed for clarity.

#### 1.2.1. Interpretations of Quantum Mechanics

There *is* a large literature about other universes, but it all seems to be anthropic, based on human interpretations of the physics and on other assumptions. For example, Hugh Everett (1971) adopted the wave interpretation of quantum mechanics as was held by Schrödinger (see Bitbol 1995). But he also felt that this picture makes sense only when observation processes are treated within the theory. So, he favored an interpretation of quantum theory by which reality is brought through observation or measurement. There is a sizable literature about the multiverse, much of it based on that interpretation of quantum-mechanics in which *observations* bring *reality*. Books and articles thereby describe universes that are a direct implication of cosmological observations; some are "parallel" universes that have identical copies of us reading this paper (De Witt et al. 1973, Valenkin 2006).

On the other hand, this paper shows that by letting equations rule without anthropic precepts one does find other universes and these have their masses, numbers, and physics specified by the equations. The paper is also based on a dozen observations ranging from nucleosynthesis to galactic clustering, but none of them affects this macro world.

### 1.2.2. Planck Time

The *standard model of our universe* starts with a "Big Bang", two commonly used words that are useful as a quick name for the model, but they indicate an explosion, an instantaneous event, which it indeed seems to be if one uses the second; all textbooks use the second. One speaks then of an event at $t = 10^{-43}$ second for the beginning, a next noted event is at $t = 10^{-32}$ s and their interval is difficult to imagine when it is expressed in seconds (Gehrels 2007d). Expressing time with the second is fine for human activities, which have their fastest reactions in $\sim 10^{-3}$ s, but not for activities of quantum fluctuations having their reactions as fast as in $\sim 10^{-43}$ s. Max Planck 1899) provided a good unit for that, the Planck Time [PT, in seconds $\sim 10^{-43}$; Eq. (3), below]. The starting time for the standard model is then at $t = 0$ PT, zero Planck Time, and that next event is now said to occur at $10^{11}$ PT because it takes as many as $(10^{-32} - 10^{-43})/ 10^{-43} \approx 10^{11}$ PT, hundred-thousand million Planck Times allowing $10^{11}$ activities.

### 1.2.3. Expansion of Intergalactic Space

Nearly all galaxies drift away from us, and this was one of the greatest discoveries of the century made in 1912 by V. M. Slipher (1875-1969) at the Lowell Observatory near Flagstaff, Arizona; he also found exceptions for our Local Group of Galaxies, which is gravitationally contracting so that there is contraction instead (Gehrels 2007d). The expansion was studied in detail by Edwin Hubble (1889-1953) at the telescopes and climate of Southern California, it is called the *Hubble expansion* or the *expansion of intergalactic space*. The latter is because the galaxies have their own gravitational regime – they do not expend, nor do we – the expansion is in the space between the galaxies. We shall refer to the *accelerated expansion* as was discovered for the epoch of some $5 \times 10^9$ years ago by two teams of Earth-based observers at large telescopes (Riess et al 1998; Perlmutter et al. 1999). This topic is in active pursuit for millions of galaxies at the greatest distances, where we see the earliest ages.

When in a thought experiment the reverse of the expansion is considered like a contraction back in time, one would end back at the beginning of our universe in zero space compressed to infinite density; that time is called $t = 0$ on the clock of the usual modeling called "the standard model". That infinite condition is called a "singularity", which has intrigued authors to write a large literature in books and encyclopedias. The next noted event, at $10^{11}$ PT, is where physicists begin to get a feel for elementary physics and particles. The following milestone comes at $t = 10^{31}$ PT with the emergence of the four nuclear forces, *i.e.,* gravitational, weak, electromagnetic, and nuclear force.

At $t = 10^{37}$ PT, $t = 10^{-6}$ sec, the theories indicate that protons and neutrons occurred, symmetrically *i.e.,* with particles and anti-particles having opposite properties. The two would largely annihilate each other; annihilation occurs when a particle and its anti-particle would meet. However, it was apparently not a total annihilation, one of the two types would happen to prevail, which is the one we now call the "particle", of which large numbers prevailed. A large amount of radiation was also produced, the photons.

Neutrinos appear at t = 1 sec., electrons 15 seconds later, and helium nuclei 1 minute later. The standard model there shows a spell of a few minutes during which the conditions were right for the assembly of the nuclei of helium, plus traces of heavier nuclei. However, the change towards less dense conditions happened fast in that expanding universe, the combination of the atomic nuclei heavier than protons could be done only during these few minutes such that only a limited number of them could be formed. Their numbers are confirmed by the amount of helium nuclei that is presently observed in the universe, and here lies a success of the standard modeling.

About following millennia we know little other than that the universe was a plasma of nuclei and electrons, with photons being scattered around, a hot, dense, noisy and energetic mass in expansion, but still too dense for light to escape. Not much was happening other than this *scattering* of light, similar to what happens inside the sun where it takes a million years for a newly made photon at the center to be bounced out to the outer levels.

Finally, at t = 380,000 years age of the universe, came the last and remarkable moment of the Big-Bang standard model, when the density was low enough for the atomic nuclei and electrons to be combined into spaciously completed atoms we know so well. The scatterers were not free anymore to scatter the photons because they were now bound in atomic configuration. A remarkable observation occurred as the spacious configuration allowed the photons of light and the waves of sound to roar out through the wide-open spaces inside the atoms.

The temperature at age 380,000 was about 3000 K. That stage can still be observed today because the expansion continued, and has even been found to accelerate, so that one speaks of the *accelerated expansion*. The boundary is all around us, far away to where it has expanded since the 380,000 years to the present age of the universe at $1.373 \times 10^{10}$ years. That boundary is now at enormous distances, observed between the stars. The effect of the expansion of the more and more tenuous material is that it cooled from 3000 K to near 3 K and it is therefore called the *3-degree-Kelvin radiation*. To third-decimal precision it is measured by spacecraft to actually be 2.728 K; this is the same 2,728 K in all directions, so that one speaks of *uniformity* and *isotropy* (Gk, isos, equal; same in all directions). But with even better precision one discovers deviations from 2.725 K in the fifth and higher-decimals. Fragmentation is then found in isolated clouds and the patterns are reminiscent of clustering of galaxies (Spergel et al. 2007).

Physicist David Wilkinson was a strong supporter of spacecraft so that a mission was named in his memory, the Wilkinson Microwave Anisotropy Probe (WMAP; anisotropy, not isotropic). WMAP draws interesting conclusions through a combination of techniques (Spergel et al. 2007). An all-sky map shows the mottling of galaxy clustering. The celestial map is further analyzed in small area segments, the deviations are plotted against the size of the areas, and certain sizes show peaks. Theory is then fitted to these peaks and this is successful because there are many observations, also by other telescopes, such that several results can be precisely obtained.

So, there is great success of the observations, and we shall use them, but they showed big problems for the Big Bang because of the extreme conditions during the early times of its standard model.

**1.2.4. Dark Energy and Dark Matter**

Another great challenge remains in astrophysics and cosmology, namely to understand the physical essence of dark matter and dark energy, which are observed but understood not at all. They are dominant in the contents of our universe so we have no choice but to include them in the discussions of this paper, and this is made attractive because the abundances are not the same at the present time as when the universe was young, so they may be clues for the origin of our universe. The percentages are from Spergel et al. (2007) in the next two paragraphs.

At the present time, the observable matter called "baryons" amounts to only 4.6%. Neutrinos have less than 1%, while 23% not-observable dark matter occurs mostly in the outer parts of galaxies. And 72% is some form of dark energy believed to be the cause of accelerating the expansion of intergalactic space; the percentages are in terms of mass or their energy equivalent.

When the universe had age 380,000 y, when photons emerged freely from matter, our universe amounted to 12% atoms, 15% photons, 10% neutrinos, or 37% baryons. Not observable but otherwise derived to be present was 63% dark matter, and a small amount of dark energy. If the dark energy had been 72%, as above, the baryons would have had 14% and dark matter 24%, so, the difference is mainly in the dark energy, as if it had been used up in the formation of our universe.

### 1.2.5. Summary of the Revised History of our Universe

After overview of Planck's basic equations, a similar equation of Chandrasekhar is folded into that regime with some restriction and calibration for its usage. Table 1 gives the range of objects inside our universe.

Before proceeding outside of our universe, the theory is checked with observations for our universe and its original stars. A search is made for other objects that might be participants in this history. No other participants are found with the possible exception of galaxies and planetesimals, which are set aside for now because of problems with their characteristics. The principal players are stars and universes.

There is one more verification of the theory before we feel confident to explore outside of our universe, and this indeed brings a set of confirmations. The linkage of Chandra's equation with those of Planck is seen, and the theory is shown to be working. The finite mass of our universe is confirmed. And, this is funny, if our Big Bang had occurred at time $t = 0$, the radius of our universe would have been that of the proton. This result may even be useful for the study of the proton.

After such encouragement to explore the entire cosmos, it is quickly found that our universe is but a small component of that cosmos. There is a large literature of multiple universes, but the multiverse of this history is known in some detail by our core equation, which provides numbers for masses and energies, its physics and therefore its principles of evolution. We know the mass and energy of the universes to be the same, *i.e.* that of our universe, and their physics are the same powerful quantum, relativity, gravity, and atomic physics of our universe.

With that, an enormous vista opens how our universe came from the large multiverse. We suddenly understand where it all came from, the evolution and intricate physics that allow us to think and study.

After such jubilation, we get back to work to discuss the death and birth of stars and everything we see in our universe. Aging is discussed. How old components of universes and even clusters of galaxies decay into the multiverse, and that the clustering of galaxies is observed today, in deep-sky observation between the stars that is called the 3-K background radiation. That clustering of galaxies survived the processes of decay and of re-birth. That is an astonishing proof of this theory, and especially for the modeling of the giant body that produced the beginning bang of our universe, that the body never became too hot to melt the galaxy-clustering characteristic away.

And then the characteristics of photons and subatomic particles are found to have survived as well. Our universe originated at a much later and further developed stage and age than the violent Big Bang would have allowed. It began with photons, protons, and neutrons some $10^{37}$ Planck times later than the Big Bang would have done, that is at $t \sim 10^{-6}$ s, a number that occurs often because it is the time when the space density is that of the proton, $10^{18}$ kg m$^{-3}$.

### 1.2.6. Comparison of Existing and Revised Histories

Such a comparison is specified for this Journal, but there is no overlap in time of existing and this revised histories, with 2 exceptions. The revised history begins $10^{37}$ Planck times later, *i.e.* at $t \sim 10^{-6}$ s of the existing clock, so that Inflation-, String-, and early-Big-Bang theories are not applicable; in fact, they had been created to solve the problems of the early stages, *i.e.*, for $t < 10^{-6}$ s.. The 2 exceptions are the interface of the revised history with Big-Bang and atomic histories around t ~ 10⁻⁶ s. And finally the revised history will state that it follows these two standard models for t > 10⁻⁶ s. The previous sections provide the context for the interface near t ~ 10⁻⁶ s.

Here ends the interdisciplinary statement of context that is specified for this Journal. We now proceed to the new history of our universe.

## THE REVISED HISTORY
### 2.1. BASIC EQUATIONS
#### 2.1.1. The Planck domain
Max Planck derived the theory of blackbody radiation with two essential constants, h and k, regarding which he said that

" "... the possibility is given to establish units for length, mass, time and temperature, which, independent of special bodies or substances, keep their meaning for all times and for all cultures, including extraterrestrial and non-human ones, and which therefore can be called 'natural measurement units' " " (Planck 1899).

That was written nearly a hundred years before the discovery of the first extraterrestrial planet, let alone an extraterrestrial culture! But this had been his goal, following his development of the basics of quantum mechanics, when he knew that at least the foundation of physics was in place and that it provides measurement units that are not anthropic such as the second, centimeter, gram, and the Celsius degree are. These units are also basic in the explanation of origins, as is demonstrated for an imaginary state at t = 0 of our universe in Sec. 2.4.

Planck found his units from dimensional analysis of the combination of cosmological constants h, G, c, and k; the four results are expressed in equations below. Values for the constants and their dimensions are h = 6.626 0693(11) x 10⁻³⁴ m² kg s⁻¹, c = 299 792 458 m s⁻¹ (in a vacuum, exact by definition), G = 6.6742(10) x 10⁻¹¹ m³ kg⁻¹ s⁻², and Boltzmann's constant k = 1.380 6505(24) x 10⁻²³ J K⁻¹ (Mohr et al. 2005). The numbers in parentheses are the estimated standard deviations. The relative standard uncertainty of G, for example, is 1.5 x 10⁻⁴.

The treatment of the present paper indicates that the Planck constant is h, not ℏ = h/2π. With ℏ, the universe's mass would be at least 30 times smaller than what we find below, the mass of a primordial star would be 15 times smaller, and the radius of the proton would be 2.4 x 10⁻¹⁶ m while its observations range between 6 and 10 x 10⁻¹⁶ m (Sec. 2.4). I have therefore changed ℏ into h where necessary.

Planck units used in this paper are

$$\text{Planck length} = (Gh/c^3)^{0.5} = 4.051\ 31(30) \times 10^{-35}\ \text{m}, \tag{1}$$

$$\text{Planck mass} = (hc/G)^{0.5} = 5.455\ 55(40) \times 10^{-8}\ \text{kg} \tag{2}$$

$$\text{Planck time} = (Gh/c^5)^{0.5} = 1.351\ 38(10) \times 10^{-43}\ \text{s}, \tag{3}$$

$$\text{Planck temperature} = (hc^5/G)^{0.5}/k = 3.551\ 37(28) \times 10^{32}\ \text{K}. \tag{4}$$

The Planck charge is also a member of the basic set and Eq. (7) is basic, while there are derived units such as the Planck density and Planck energy in Eqs. (8) and (9). The entire set is the Planck domain.

#### 2.1.2. The "Universal Planck Mass", M(α)
More atomic physics may be added to the Planck domain as a result of the theory of structure, composition and source of energy for stars by Chandrasekhar; such calculations are complex and involve a variety of physical laws. He had developed that discipline with detailed laws such as of Stefan and Boltzmann, relating pressure and temperature at various depths inside the star. The laws brought the Planck constant h, the velocity of light c, Newton's gravitational constant G, and the mass of the proton H (Chandrasekhar 1951, pp. 599-605). For the total stellar mass, M, he found,

$$M \approx (hc/G)^{1.5} H^{-2}. \tag{5}$$

He had also found a generalization that yields cosmic masses,

$$M(\alpha) = (hc/G)^{\alpha} H^{1-2\alpha}, \tag{6}$$

for positive exponents $\alpha$, which identify the type of object, such as $\alpha = 2.00$ for our universe as well as the above $\alpha = 1.50$ for stars. The equation can also obtained by dimensional analysis, but his derivation justifies the usage of h, c, G, and H, also for the Planck domain.

## 2.2. RESTRICTIONS OF M($\alpha$)
This paper puts two limitations on membership of M($\alpha$), to begin with the following section. Equation (6) is simplified - and calibrated at the same time - in Sec. 2.2.2. Section 2.2.3 has a summary Table, which will be the mainstay for the discussions.

### 2.2.1. Restriction to Primordial Baryonic Objects
This paper deals with baryonic masses, *i.e.* consisting of observable matter only and that is only 4.6% of our universe. The compositions other than 4.6% are in Sec. 1.2.4.

The treatment of this paper is only for origins of primordial masses. In the case of stars, for example, the usage of M($\alpha$) is limited to original matter consisting primarily of hydrogen and helium, rather than of the later compositions in subsequent stars that have increased abundance of heavier elements. For the universe, the usage is limited to its very beginning until 380,000 y.

### 2.2.2. Calibration with the Proton Mass
A simplification of Eq. (6) is made by expressing the masses in terms of the *universal mass unit* of the proton mass, 1.672 621 71(29) x $10^{-27}$ kg, such that H = 1, and

$$M(\alpha) = (hc/G)^{\alpha}, \tag{7}$$

in proton masses; at $\alpha = 0$, M($\alpha$) = 1 proton mass.

Note the similarity of $(hc/G)^{\alpha}$ to the Planck mass $(hc/G)^{0.5}$, such that Eq. (7) may be considered to be the *universal* Planck mass. The definition of the Planck mass – what it is used for – has been rather unclear until now. It had been theorized in an original phase of matter, an impossibly high Planck-density phase. The Planck mass was then imagined to be compressed within one cubic Planck length such that most of its components, but not all, can interact at velocity c, which is a Planck length in a Planck. These concepts have served to define the Planck mass, while we now replace that with its role in the mass scaling of the cosmos. This is demonstrated in Sec. 2.4 by using $c^5/hG^2$, which is the

$$\text{Planck density} = 8.2044(25) \times 10^{95} \text{ kg m}^{-3}. \tag{8}$$

It is also essential to remind ourselves that when the mass of M($\alpha$) is known, its energy is also known because of their equivalence, $E = mc^2$, and there is the unit of the

$$\text{Planck energy} = (hc^5/G)^{0.5} = 4.903\ 20(36) \times 10^9 \text{ Joule}. \tag{9}$$

We are now ready to overview all members of M(α) collected in the Table.

### 2.2.3. Overview Table

Table 1 presents data first in the universal unit of the proton mass, but then also in solar masses (s.m.) or kilograms in order to get a feel for the objects in our anthropic world. These are representative proton masses – it does not say that there were $10^{78}$ protons in the primordial universe. The class of those objects is next, and the last column gives the α-values for the observations in Secs. 2.3.1 and 2.3.2. The observed α-values of the bottom two lines, laboratory values, are the same as predicted to high precision. The Table shows quantization, such that Eq. (7) may be written as

$$M_N = (hc/G)^{0.5N}, \qquad (10)$$

with values of N in the second column of the Table. The third column is computed with Eq. (10), the fourth with Eq. (6).

**Table 1. Computed and Observed Masses**

| α | N | Computed proton masses | units shown | Type of Object in M(α) | Observed α |
|---|---|---|---|---|---|
| 2.00 | 4 | 1.131 79(35) x $10^{78}$ | 9.5172 x $10^{20}$ s. m. | Primordial Universe | 1.998-2.008 |
| 1.50 | 3 | 3.469 96(79) x $10^{58}$ | 29.179 s. m. | Primordial stars | 1.49-(1.53) |
| 0.50 | 1 | 3.261 68(25) x $10^{19}$ | 5.455 55(40) x $10^{-8}$ kg | Planck mass | 0.50 |
| 0.00 | 0 | 1 | 1.672 621 71(29) x $10^{-27}$ kg | Proton | 0.00 |

s.m. = solar masses; the brackets have estimated standard deviation

### 2.3. THE FIRST COMPARISONS WITH OBSERVATION

In order to verify the reality of M(α), comparison must be made with observations for the earliest configurations of our universe and stars. And Sec. 2.3.3 then has a search for any other objects that might have to be included in the treatment of M(α).

### 2.3.1. The Primordial Universe

The title means that the mass of the universe is considered at its beginning, even though that beginning is not precisely defined, it may be somewhere between t ~ $10^{-10}$ seconds and a few minutes (t being on the clock of the universe's standard model). Again, we are dealing here with the 4.6% baryons – dark matter and dark energy are not considered in this paper until Sec. 2.6.5 and then in a different context. For comparison with observations, there is a most appropriate analysis of nucleosynthesis at t ~ 1 min for deuterium, helium-3, helium-4, and lithium-7 (Copi et al. 1995). That result is for a density, and to make comparison with a mass, one multiplies of course with a volume, but which is the appropriate volume?

There is the apparent volume of the presently observable universe, which takes all observational effects and their corrections into account for an expanded mass (Riazuelo et al. 2004, Lineweaver et al. 2005), but such effects did not occur for an early mass and the corrections are therefore not applicable. The comparison with the observed densities for t ~ 1 min should however take the expansion itself into account because the primordial density was observed in modern times.. That is the volume of a spherical universe having radius of curvature of 1.373 x $10^{10}$ lightyears, consistent with the expansion-age determination for our universe of 1.373 (+.013, -017) x $10^{10}$ years (Spergel et al. 2007). The baryon densities derived from the nucleosynthesis lie between 1.7 and 4.1 x $10^{-28}$ kg m$^{-3}$ (Copi et al. 1995), for which the above volume-derivation method yields between 9.27 x $10^{77}$ and 2.24 x $10^{78}$ masses expressed in terms of proton masses; in terms of $\alpha$ that is between 1.998 and 2.008.

The difference between observed and theoretical values of $\alpha$ is primarily due to the above spread in the density determinations, while the effect of the above spread in radius for 1.373 x $10^{10}$ ly is less, which appears to confirm the method of determining the volume of the universe. Another confirmation is in Sec. 2.4, where this mass for our universe yields a determination of the size of the proton. Yet another confirmation is to follow the standard modeling, using the same method but now to determine the age for our universe when it had the proton density of Eq. (13) and this yields $10^{-5}$ sec, in good agreement with the standard-model derivation from other data, at $10^{-6}$ s.

### 2.3.2. Primordial Stars

This title is defined at the end of Sec. 2.2.1. First, it is noted that physical laws for pressure and temperature within stellar interiors establish that the theoretical value of exponent $\alpha$ is exactly 1.50 (Chandrasekhar 1951, Carr et al. 1979). The selection of observations for the present comparison is from stars that settled towards "early" spectral type O, and their values lie near 10 solar masses ($\alpha = 1.49$) when they have not yet accumulated most of their final stellar mass (Stahler et al. 2000, which is a review chapter). The star settles into equilibrium in a steady state, but it has a short life and ends as a supernova with two shockwaves, the first of radiation, followed by a slower one of matter. The latter delivers atomic nuclei to the interstellar medium, of atomic weight higher than those of the original hydrogen and helium; heavier nuclei had been formed.

Reports appear in the literature of much more massive stars, but they are either resolved as stellar clusters, or they consist of accreted masses or of highly unstable and shedding mass, or they have much heavier than hydrogen-and-helium composition (Bonnell et al. 2004). An extreme of 500 solar masses (Bromm 2006) is represented in Table 1 as 1.53, but with doubts that it fits the criteria of Sec. 2.2.1, or that it would be included in the derivation of Eq. (5). Anyway, it is noted that 500, which is a factor of as much as $500/29.179 \approx 17$ larger than predicted, is only an exponential $\Delta\alpha = 0.03$ off from predicted, which is small compared to the interval between categories in the Table of a factor of $10^{19}$ or $\Delta\alpha = 0.50$.

Incidentally, "solar mass" in Table 1 merely indicates a unit of 1.9891 x $10^{30}$ kg, rather than a solar-type star; the number of early-type stars considered here is not 9.5 x $10^{20}$, but 3.3 x $10^{19}$, 29.179 times smaller.

Are there any other participants in M($\alpha$), other than a universe with stars?

### 2.3.3. A Search for Completion

A search was made for additional members of M($\alpha$) constrained by Sec. 2.2.1. A variety of spiral galaxies had been observed at the 21-cm hydrogen-line, showing on average 5 ($\pm 4$ standard dev.) x $10^{10}$ solar masses (Cox 2000). One should perhaps select the upper limit to allow for dissipation of mass and energy through collisions, but there is also accretion from dwarf galaxies (Schweizer 2000). The upper limit for galaxies is near $10^{12}$ solar masses (Carr et al. 1979). Observations have also been made of young galaxies at great distance; this was actually on multiple galaxies occupying a single dark halo, and $10^{11}$-$10^{12}$ solar masses were found (Ouchi et al. 2005). The observed range for all galaxies is $10^7$-$10^{12}$ solar masses, which represents $\alpha = 1.64$-1.77, but that includes subsequent development (Schweizer 2000), while we are interested in the youngest galaxies. In summary, the range of $\alpha$ in the last column of Table 1 might then have been 1.72-1.77, representing the range of $10^{10}$-$10^{12}$ solar masses.

However, I decided not to include galaxies in Table 1, for the following five reasons.
- Even though the fit of galaxies to the prediction at $\alpha = 1.75$ is good relative to its large quantization interval that would have been $\Delta\alpha = 0.25$, it is not as good as the fits for stars and the universe.
- There are questions whether the galaxies formed at $\alpha = 1.75$ or that they subsequently accreted towards that size.
- There may be a problem that births of stars and galaxies are intertwined; the WMAP spacecraft finds evidence of early star birth (Spergel et al. 2007).
- With all exponents of Eqs. (1) – (4) being 0.50, nature seems to point at such quantization, instead of $\Delta\alpha = 0.25$ if galaxies were included.
- The same conclusion in favor of 0.50 is near the bottom of Table 1, with the Planck mass at $\Delta\alpha = 0.50$ separation from the proton mass and the same is seen between the stars and universe.

It therefore seems prudent to leave the topic until additional observations and interpretations of primordial galaxies and stars are made. However, it will be noted in Sec. 2.6.5 how indicative galaxies are in the 3-K background observations, such that they prove the validity of our modeling. The decision to add galaxies to Table 1 can always be made later, and $\Delta\alpha$ will then be 0.25.

The stars in open and globular clusters should not be considered for $M(\alpha)$ membership, because they consist mostly of subsequent atomic nuclei, such that they show spectra usually much later than type O.

At $\alpha = 1.00$, planetesimals of rocks and soil at approximately 1-km radius do not have the type of material specified in Sec. 2.2.1, even though they appear to be primordial in the solar system (Alfvén et al. 1976, Kleczek 1976). This is also an open problem for further consideration.

No other members were found for $M(\alpha)$; Table 1 has all that are in our universe having membership defined in Sec. 2.2.1. Early-type stars therefore are the smallest primordial members of the universe that consist primarily of hydrogen and helium and are stable enough for application of $M(\alpha)$.

An independent verification of $M(\alpha)$, supporting worthwhile theory, is next.

## 2.4. VERIFICATION WITH PROTON-RADIUS OBSERVATIONS

This section began with wondering, "by back-tracking the expansion of intergalactic space, some modeling has zero size and volume for our universe at $t = 0$, which is a singularity, nonsense actually, so what would the Planck domain have - what would be the size of our universe at Planck density?"

Here are three exercises to find the answer. First, the constant factor F between steps of $\Delta\alpha = 0.50$ in the Table is seen in the number of proton masses for the Planck mass, which is the same number as for primordial stars generated in our universe,

$$F = 3.261\ 68(25) \times 10^{19}. \quad (11)$$

The ratio of the universe's mass and the Planck mass has the third power of F, but the third root of that is then taken for the length ratio from that volume ratio, coming back to F. Thus we obtain a size parameter for the universe at Planck density of Eq. (8) from the product of F and Planck length in Eq. (1). That is however the size of a rib of a cube (mentioned for the explanation of the Planck mass in Sec. 2.2.2), while for the radius of a spherical volume for the universe one divides by the cube root of $4\pi/3$ to obtain $R' = 8.1974(9) \times 10^{-16}$ m.

A basic exercise, but without demonstration of the cube in the explanation of the Planck mass, is to divide the mass of the universe in Table 1 by the Planck density of Eq. (8), and obtain radius R' again.

A third and more precise derivation is made by realizing that the *formulae* for the universe's mass, $(hc/G)^2 H^{-3}$ [H also in kg], divided by the one for Planck density, $c^5/hG^2$, yield the volume of $h^3 c^{-3} H^{-3}$. The low-precision gravity term, G, takes no longer part of course, and that increases the precision of the derivation. After rib-radius conversion again, the radius of the universe, if it would ever have been at the unlikely Planck density, would have been,

$$R = 8.197\ 3725(20) \times 10^{-16} \text{ m}, \qquad (12)$$

with the precision depending only on those of h and H, since c is exact and G is no longer involved.

This last exercise provided a derivation of a radius for our universe at Planck density, and it looks surprisingly alike the size of the proton. How does Eq. (12) compare with observations of the proton size? For a comparison with charge-radii observations, a straight average of the radius obtained by various teams (Karshenboim 2000) gives 8.2 (±.3) x $10^{-16}$ m for six observations, of which however there is one as far off as 6.4 x $10^{-16}$ m, while five of them are between 8.09 and 8.90 x $10^{-16}$ m. Two other observations yield 8.05 (±.11) and 8.62 (±.12) x $10^{-16}$ m (Berkeland et al. 1995). It is seen from the high precisions of widely different results that the determination of Eq. (12) could only be for an *equivalent* proton radius, rather than claiming that the proton is a sphere. The word "equivalent" is then for a hypothetical spherical shape of the proton. The proton has for a long time been considered a *fuzzy* sphere having radii between 6 and 10 x $10^{-16}$ m. A better interpretation for the above precise measurements resulting in a wide variation of proton size, is that its shape is time-dependent, perhaps due to internal quark motion (Berkeland et al. 1995).

Another choice instead of H was tried for the computation of Eq. (12), namely the mass of the $^1$H atom, which seems a small increase but the result is grossly off Eq. (12), at R = 8.192 9019 x $10^{-16}$ m. It is seen that for any value of H larger than that of the proton mass, R would be smaller, and vice versa because of the inverse proportionally in Eqs. (5) and (6). A smaller size for a larger mass and vice versa? Does Eq. (12) converge on the proton radius as some absolute value? It is also remarkable that the size of the proton has the same number of Planck lengths as the mass-quantization constant F of Eq. (11).

The equivalent density of the proton, assuming uniformity, follows from Eq. (12) and the proton mass,

$$\text{equivalent proton density} = 7.249\ 1169\ (54) \times 10^{17} \text{ kg m}^{-3}. \qquad (13)$$

The above radius relation is steep, such that this exercise also serves as a confirmation for the mass of our universe and for the internal consistency of the theory. If the mass of our universe would have been for example a factor of 2 larger ($\alpha = 2.008$), it would have yielded the radius at 1.02 x $10^{-15}$ m, which is out of the question when compared to R and its precision. This result indicates *fine-tuning* for our universe with high resolution compared to nature's large quantization factor of F ~ $10^{19}$ shown between the steps in Table 1 [Eq, (11)].

The curiosity question asked at the start of this section brought five discoveries. It shows the linkage of $(hc/G)^\alpha$ with the first three Planck units, Eqs. (1) – (3). This thereby confirms the theory and its finite mass of our universe, and it yields a theoretical radius of the proton, which is the radius of our universe if it ever were at Planck density. These confirmations and results encourage us to proceed with this theory towards the multiverse.

## 2.5. OBSERVATIONAL EVIDENCE THAT THERE *IS* A SPECIFIC MULTIVERSE

Application of M($\alpha$) beyond $\alpha = 2.00$ *must* be explored for 6 reasons.

• The Supply Problem: where did our universe's observed energy equivalent to $10^{21}$ solar masses come from? The problem apparently is not solved in inflation theory either. I asked a distinguished inflation theorist, John Heise (pers. comm.., 2008), about that and he replied,

> "" In Inflation theories, the idea is that the matter in the universe comes from a conversion of the inflation energy at the end of the inflation epoch, but this is precisely where all inflation theories are in difficulty. They cannot explain this in detail. Up to now no complete self-consistent inflation theory exists for that matter. It is still only an intriguing idea without final theoretical proof, but somehow everybody thinks that will come eventually.""

• A mystery with assuming a sole universe is where our physics could have come from. How could something so intricate have developed in the beginning of our universe when the techniques of evolution were still primitive and all stages lasting short times? In the early times of the standard model, the density changes so fast that there does not seem to be enough time for the evolution of our intricate physics.

• The First Uniformity Problem: uniformity is observed to third-decimal precision in the 3-K background observations by WMAP and others (Spergel et al. 2007); the temperature of 1.725 K is observed in all directions. How does that uniformity come about? It has been assumed that the universe would have had an exceedingly small size such that all components would have interacted, but could there be another cause for the uniformity (Sec. 1.2.3)?

• The Second Uniformity Problem: how could the fifth decimal of the 3-K background show appreciable non-uniformity with variations on a scale of galactic clustering (Spergel et al. 2007)?

• Quantum theory and associated physics produced $(hc/G)^{0.5N}$ yielding a mass at any value of N, as is shown in Sec. 2.2.3, and M($\alpha$) *is open* to values of $\alpha > 2.00$, $N > 4$.

• It follows from the Table that the larger values, of $\alpha = 2.50, 3.00\ldots$ et cetera mean that there is a specific multiverse. This is further discussed in the remainder of this section, while the 6 points and their questions are answered in the following Sec. 2.6.

The numbering of the alphas in M($\alpha$) is an anthropic curiosity because we think of our universe as having $\alpha = 2.00$, while another culture in another universe will do the same for its own universe. Imagine another culture elsewhere, with its intelligent beings seeing a hierarchy as we do, within and outside of their universe. They too start with their symbols for h, c, G, and H, which are measured in their laboratories, and they consider that their "$\alpha$ values" are from 0 for the proton mass to 2.00 for their "universe", and onward out into "the multiverse". Their universe is then imagined by them as we do for ours, as a member of their Local Group of $3.3 \times 10^{19}$ universes at $\alpha = 2.50$, which is a member of $3.3 \times 10^{19}$ Clusters at $\alpha = 3.00$, et cetera, following their factor F [Eq. (7)]. While no one may ever know the ultimate largest number of the $\alpha$-values, there appears to be a finitude to the multiverse, which seems more beautiful and therefore more truthful than an open infinity (Chandrasekhar, 1987).

The physical truth is that more universes are always observable at greater distance, as is seen for galaxies within our universe. The truth is also that M($\alpha$) has a quantization; with numerical value of the factor F in Eq. (7). It is seen in the Table between proton and Planck masses, between the stars and the universe, while there is a multiple of F between Planck mass and the stars. It also occurs between Planck length and proton size, which seems to confirm the factor as a basic effect.

A different approach to the understanding of quantization is by studies of what limits each category. Carr and Rees (1979) have followed the practical approach for stars, using theory of star formation. They derived the limits to mass and energy in an accreting cloud of hydrogen and helium in order to make an object that survives as a star.

## 2.6. THE HISTORY OF OUR UNIVERSE
Section 2.6.1 details the ending of the history, and this brought the need to study the multiverse as an evolutionary system, in Sec. 2.6.2. The beginning of the history as the birth of our universe is in Sec. 2.6.3. The last section has a different view, which brings a confirmation and a study of dark energy.

### 2.6.1. Decay of our Universe
Everything in our universe ages and decays. Even the proton, basic part of every atomic nucleus, may have a limited life. Andrei Sakharov (1965) computed that its half-life is $10^{50}$ years or larger, while its increasingly difficult observational verification stands at $10^{35}$ y. This is an important result because it implies that the cosmos would end unless there is renewal; this cosmos would be a unique event. Sub-atomic particles are therefore fundamental in the history of our universe.

Photons are also fundamental as wave phenomena (Lamb, 1995). They emerge from stars, supernovae, gamma-ray bursters and all other sources of radiation; their aging is in terms of radiating out into space and thereby cooling to near 0 K. The sub-atomic particles and old cold photons move out on the accelerated expansion as well as whole galaxies and clusters of galaxies and whatever other debris such as of old and remnant stars. The percentages of dark matter and dark energy must be included because they occur in our universe; they are in Sec. 1.2.4.

The discovery of *accelerated* expansion by the teams of Riess et al. (1998) and Perlmutter et al. (1999) is essential for this history, for the debris to eventually be captured in the inter-universal medium (IUM). Within a large multiverse, the debris will eventually encounter debris objects from other universe that are expanding into other directions. In three dimensions, the result over long timescales is bound to be some mixing by the old and decayed components of universes. Any expansion of the multiverse itself must be a secondary effect to be modeled with the effects of mutual gravity and radiation pressure. The modeling that seems needed here is a Monte-Carlo type simulation in a large box in the multiverse.

Galaxies travel individually on the expansion at first, but they eventually also get caught in the IUM. Clusters of galaxies apparently survive the entire procedure as they are recognized in the present 3-K observations (the fourth point in Sec. 2.5).

Information from the inter-stellar medium (ISM) is useful because some of the same processes are bound to happen in the IUM over cosmological scales of space and time. There again is a continuous supply, but its composition is now totally different, it is of the above inert and decayed debris instead of atomic and molecularly active ISM material. It has uniformly mixed composition because it is fed by input from various universes but its space density will be locally uneven with huge clouds that include galaxies. Nothing stands still in the cosmos - the large clouds continue to grow by sweeping up of the material during their motion through space, like planets do during their formation. Self-gravitation will become active, speeding the contraction of the cloud towards making a new universe with increasing gravitational cross-section. It is no longer active material of hydrogen and other atoms; instead, it is *energy-seeking* material of cold and old debris. The growing proto-universe can therefore increase in mass without getting as hot as a proto-star would have become, while the gravitational energy of the compaction is used to re-energize photons, and to re-energize and re-constitute the atomic components into regular protons, neutrons, and other particles (this will be taken up in Sec.2.6.3). Eventually the IUM clouds complete their growth gravitationally as protostars do from interstellar matter, but now on cosmological scales.

The IUM scales of number and size are larger by a factor of ~$10^{19}$, the quantization Factor F of Eq. (11). We may make a coarse estimate of the cosmological *time* scale as follows. An *upper* limit comes from Sakharov's >$10^{50}$ years. If renewal is obtained appreciably sooner than in $10^{50}$ years, the proton would not have to be re-constituted, but it still may need to be re-energized. The more basic sub-atomic particles would not need re-constitution. A trial-and-error evolutionary multiverse (next section), apparently evolved protons with half-life that long. The *lower* limit is known from what is observed for stars, a half life on the order of $10^{11}$ y for the slowest. The time scale of universes within the multiverse appears to be somewhere in between the two limits, very roughly at $\sim 10^{30}$ y, with a factor $10^{30}/10^{11} \sim 10^{19}$ as in Eq. (11).

The question arises if we might not see in our present universe some of the debris of other old universes. The first reply might be that whatever caused the acceleration of the expansion, mentioned above, would have a protectively repulsive effect, especially on galaxies. Also, that our Local Group of galaxies is contracting. However, Basu (2006) has been searching the literature persistently for galaxies that have blue-shifted spectra, Basu's number of blueshifts is small (some 141 cases from searching the literature ~20 years) among the huge number of observed redshifts (millions), indicating that the mixing is quite limited over our universe's time scale of $10^{10}$ years. Encouragement of Basu's pioneering and additional search seem advised.

## 2.6.2. Evolution in the Multiverse

For a more detailed history, the principles of evolution are used from a text that shows at least 13 of them for the inorganic domain of our universe (Chapter 7 of Gehrels 2007d). Nature searches randomly for continuation, in trials depending on natural selection, with undirected slightly different characteristics – will they yield survival, or is an error made so as not to survive? Evolution thereby evolves itself in ever developing more complex forms or species, which usually means that they are more capable. This can be observed in nature around us and within ourselves. This is in fact how the most advanced tools of organic evolution emerged from the organic ones.

Regarding the essence of evolution further, we saw that the result depends on the environment, but the environment is also modified, interactively. Much of evolution is unpredictable, but there seems to be an overall trend towards greater complexity, which usually brings greater capability such that even evolution itself evolves.

One can confine the modeling by considering a relatively small volume within the multiverse, but still with a multitude of universes, and considering this as a closed system in which mass is conserved and there is interaction. The trial-and-error evolution then has universes originating within that volume from the inter-universal medium (IUM) as is specified by the $M(\alpha)$ equation [Eq. (7)]. Failures vanish back into the IUM. A universe may happen to originate with characteristics that deviate too much, such as the wrong mass – survival occurs only near those observed in our universe. How near?

Our universe's characteristics are exceedingly tight. Fred Hoyle's name is attached to the extremely low probability for the fine-tuning of the nuclear transitions within stars. He religiously pointed out that the selections and combinations could not have occurred if the physical constants of the elements would have been even slightly different. But now we have seen the physics of our universe in unified operation of quantum, relativity, gravity, and atomic physics with the $M(\alpha)$ equation precisely calibrated on the proton. If our universe resulted from and is decaying into the IUM, the IUM must have that physics. The IUM has complete homogeneity through mixing of debris eventually from a large number of universes. All universes sprouting from and decaying into that medium have that same physics. Hoyle's problem seems solved because the continuing evolution within the multiverse produces finely tuned universes to begin with. The multiverse has long times and the IUM environment is controlled by h, c, G, and H.

The universes have to have near-critical mass or they cannot survive the evolution, they would either collapse or expand rapidly into nonexistence. This is what is meant when WMAP states that our universe is "nearly flat", near critical mass (Spergel et al., 2007). It is a requirement for any model of universes.

Galaxies are important components in evolution. Their collisions appear to be rather elastic, often resulting in a new galaxy (Schweizer, 2000); they bring possibilities tried in interactive but random "trial-and-error" searching for what might survive.

Further examples of the evolution are the following. The result depends on the environment, while the environment is also modified, interactively; natural selection is seen, which is the key to evolution. Much of evolution is unpredictable, but there seems to be an overall trend towards greater complexity, which usually brings greater capability so that evolution itself evolves. In the end, the surviving universes obtained the unified quantum, relativity, gravity, and particle physics in a most advanced form for this epoch, but the process never stops.

**2.6.3. The Beginning of Our Universe**
There are observations supporting that this history of our universe and its timing are correct, and it is noted that some of them show also that the temperature stayed low enough for the survival of the characteristics for components from defunct universes.
• The proto-universal cloud continued to collect material, apparently up to the baryon limit for a universe, which is equivalent to some $10^{21}$ solar masses, but the miracle of evolution in the multiverse is that this coincided with proton density near its center. This should be verified in detailed modeling – the problem should be posed for $10^{18}$ kg m$^{-3}$ as a must at the center and possibly also being the density of all the material outward (for dark matter as well), and then the simple question if the total mass can be equivalent to $10^{21}$ solar masses.

There is the peculiar fact that only 4.6% of the mass of our universe is baryonic, visible matter. Why it is so low has been a persistent query, but now it may be seen from simple geometry as follows. As the mass of the IUM cloud grows towards the equivalent of $10^{21}$ solar masses, it is of uniform composition because the debris of many universes is mixed in together, having all the above components of photons, protons, neutrons, old stars, galaxies, and dark mass and dark energy. The latter has at this time still the 72% of Sec. 6.1.

The density in such a gravitational cloud varies of course, with the highest in the center and falling off outwards. However, in the above it was also noted that "a uniform sphere of density σ," worked well in the first line of Table 2, indicating that "$10^{18}$ kg m$^{-3}$ as a must at the center and possibly also being the density of all the material outward" is sustained. The conclusion is that the falling off outwards of the clouds density is tempered. Proof of this was found also at t = 380,000 y, in the above.

Anyway, when about 4.6% of the total mass near the center reaches proton density of $10^{18}$ kg m$^{-3}$, the subatomic particles of the proton and the dark energy are forcefully pressed together. The enormous gravity effects near the center of a sphere with the mass of our universe are used for the re-constitution directly. The particles are thereby re-energized and re-constituted by dark energy as well as by the gravity itself to make newly constituted protons and neutrons. The assumption of such activity and effect by dark energy is based on the interpretation that the *acceleration* of the expansion is due to the activity and effect of dark energy (Sec. 6.1). Perhaps one could derive the 4.6% numerically – the point here is merely to answer the question why it is such a low percentage.

There are more observations that this history of our universe is correct:
• Outside of the 4.6% volume of baryonic mass, the re-energizing and re-constitution could not take place because of decreased density; in those dominantly larger outer reaches of the sphere, the density was not high enough for recombination. The 63% dark mass in Sec. 6.1 remained, alike it does in the outer regions of galaxies. The dark mass played the role in the process of allowing 4.6% within 100%, and the multiverse evolution therefore evolved dark matter. Dark mass is closely related to the debris from old universes.
• The dark energy is used up. This is seen by comparing the paragraphs for the present and 380,000 years in Sec. 6.1, for 73% and ~0%, respectively. The multiverse evolution did evolve dark energy to make the system survive.
• The standard models do have t ~ $10^{-6}$ s as the time and density for photon as well as proton generation. However, the re-energizing of photons appeared a little earlier because it is a simpler process than re-constituting. It had only a short time to burst out, to escape rather explosively to kinetically overcome the gravity, before the remainder of the plateau got to be re-constituted as well as re-energized. After this we know of high opacities until age 380,000 years. So, a short flash of radiation appeared, apparently vigorous enough to provide the radiation signature observed by WMAP with a wider curvature than that of the 3-K radiation (Spergel et al., 2007).
• Another proof is that the characteristic of the galactic clustering input from debris of old universes is preserved and appears to be generally observed by WMAP in our new universe. This characteristic might however have prevailed lower than $10^{18}$ kg m$^{-3}$, perhaps within the dark matter.

The recombination inside the 4.6% central volume apparently caused a rather explosive event, of which the WMAP signature is the sudden appearance of a 0.6- AU sized baryonic mass equivalent to $10^{21}$ solar masses, where there was only dark sub-atomic matter before. The explosion was enhanced by the sudden reversal from in-fall accretion to outward expansion driven by further particle development. It generally was the time of proton density $10^{18}$ kg m$^{-3}$. Spacecraft observations indicate an energetic event for the beginning of our universe (Spergel et al., 2007), so it seems appropriate to speak of a "Proton Bang", in deference to the label of "Big Bang" before. The Spergel et al. observation is presently being explained as a confirmation of inflation theory - instead, I ask here that the sudden appearance of a baryonic object measuring ~0.6 AU be considered to have occurred at $t \sim 10^{-6}$ s.

**2.6.4. Dark Energy causing the Expansion of Intergalactic Space**
A special confirmation comes from the Schwarzschild radius, which gives an upper limit of a mass and radius combination below which radiation cannot escape, and the object may become a black hole. Karl Schwarzschild wrote it in 1916, while dying of pemphigus, as a part of his detailed derivation of one of Einstein's approximate equations, but the part regarding the Big Bang has been ignored. The arithmetic is simple because for light to escape, its kinetic energy must be greater than the local gravitational potential. For the velocity of light, c, it follows that the limiting radius of the object is $R_S = 2GM/c^2$. Table 2 shows the comparison of $R_S$ with radius $R = (3M/4\pi\sigma)^{1/3}$ for a uniform sphere with density σ.

**Table 2. Radii and Schwarzschild Radii**

| $R/R_S$ | t | σ | R(ly) |
|---|---|---|---|
| 1 | 380,000 y | $10^{-20}$ | $10^{7}$ |
| $10^{-14}$ | $10^{-6}$ s | $10^{18}$ | $10^{-5}$ |
| $10^{-40}$ | 0 | $10^{96}$ | $10^{-31}$ |

The first line applies when the universe's radiation is known to escape at $t \sim 380,000$ years. Schwarzschild's modeling looks good because standard theories predict the density to be $\sim 10^{-19}$ kg m$^{-3}$ at that time.

The second line is for the above $t \sim 10^{-6}$ s, using proton density of $10^{18}$ kg m$^{-3}$ in R. Because $R/R_S = 10^{-14}$, the Schwarzschild radius strongly prohibit the beginning of our universe. However, the size of the universe, $10^{-5}$ lightyears = 0.63 astronomical units AU, is comparable even though much larger than the sun, inside which it takes a million years for a photon to escape from its center. For the dark matter and un-energized debris in the outer regions of the universe, as was found above, that duration would be appreciable less, such as 380,000 years. The mechanism would then be a combination of radiative scattering for sure, and possibly also of dark energy because it played a similar role in the acceleration of the expansion, this time to hold the body up against black-hole collapse. The proof is two-fold. First, the above paragraphs that are marked with • show that the dark energy has been used up at the time of formation of our universe. Second, the body is not merely held up, it expands; the expansion of our universe has begun. The trial-and-error evolution in the multiverse evolved dark energy to make the recycling of universes possible, in the acceleration of expansion, and to produce expansion to begin with.

In the third line, the Planck density of $10^{96}$ kg m$^{-3}$, if our universe were ever at t = 0, is used for obtaining R and thereby R/R$_S$. The Big-Bang idea came from thinking the expansion backwards to t = 0, that is with all the mass of our universe in nearly-zero space, which is a grossly unrealistic condition. And this time, neither radiation pressure nor dark energy is available as yet to save the modeling. The Table shows that the thought-experiment's reversal ought to have been halted well before that troublesome Big-Bang time, namely at $10^{-6}$ s.

After t ~ $10^{-6}$ s, equal to t = 0 on the new clock, our understanding of the physical evolution appears to be back on track of the standard models for our universe and for particle physics. Our universe began with photons, protons, and neutrons, ready to go on the path described by the old standard models.

**2.7. CONCLUSIONS**
During the years of searching for this history of our universe, new insights and observations have invariably brought progress, and this process has not stopped as yet. New ideas keep coming. These are perhaps indications of truth for the model, as are its common sense, internal consistency, and beauty (Chandrasekhar 1987). If so, one can turn the reasoning around, assuming this history is nearly correct and thereby making predictions for new observations. This is much more difficult and it depends on the researcher and facilities, but some may be derived from the following conclusions.
1. The M(α) equation has proton and Planck masses at its foundation, it is connected to the Planck domain and is recognized as a *universal* Planck mass.
2. The equation, when restricted to primordial baryonic masses, compares well with observations for primordial stars and our primordial universe.
3. The M(α) equation uses quantum, relativity, gravity, and atomic physics together in a unified manner.
4. The interpretation of quantum mechanics that is used here is without dependence on observation; this was found for the macro cosmos.
5. Observations of the proton size have been verified, and that procedure of verification confirms that α = 2.00 gives the mass of our universe.
6. The equivalent spherical radius of the proton is 8.197 3725(20) x $10^{-16}$ m.
7. The treatment of this paper indicates that the Planck constant is h, not ħ = h/2π (Sec. 2.1.1).
8. It also indicates that the cosmological constants h, c, G, and H are constant over a considerable range of location and ~$10^{30}$ years. It is however an evolving universe such that the constants also will change albeit imperceptibly slowly.
9. Six reasons are given why there is a multiverse with specific mass and physics for each universe.
10. The multiverse is a quantized hierarchy of exponentially increasing numbers of universes.
11. There is a quantization factor between masses, F = 3.3 x $10^{19}$, and it is the same between the Planck length and proton size (Eq. 12).
12. All universes must have near-critical mass in order to survive.
13. The description includes that of the origin of our physics because the inter-universal medium must have it, and therefore all universes as well.
14. Fred Hoyle's fine-tuning of the nuclear transitions within stars is explained because the continuing trial-and-error evolution within the multiverse produces finely tuned universes to begin with.
15. The cosmological foundation of our world and its physics is in trial-and-error evolution within its hierarchy of universes.
16. Universes apparently decay into sub-atomic particles and cold photons over cosmological time scales and at near-absolute-zero temperatures.
17. The accelerated expansion of intergalactic space brings mixing of universes over long cosmological times.
18. The cosmological time scale is near ~$10^{30}$ years; generally, the above factor of $10^{19}$ applies.
19. The inter-universal medium thereby consists of all possible components, including galaxies that are gravitationally held together, and clusters of galaxies. Protons, neutrons, electrons, dark matter, and dark energy are included as well as old stars and stellar remnants.
20. The Schwarzschild radius prohibits any stages earlier than t ~ $10^{-6}$ s, and perhaps even that. The beginning of our universe may then have been at some time between t ~ $10^{-6}$ and 380,000 y.
21. The description is consistent with having dark matter and dark energy.
22. The percentages of baryons, dark matter, and dark energy are determined by the spherical geometry of the proto-universe during its highest density.
23. The classical wondering why the baryons have only 4.6% of the composition follows readily from that spherical geometry.
24. The subatomic matter may be re-configured into protons, neutrons and radiation in a "Proton Bang" exploding into the beginning of a universe near $10^{18}$ kg m$^{-3}$ at t ~ $10^{-6}$ s (on the clock of the present standard model).
25. This event might then be interpreted, instead of as inflation, as the sudden appearance of a baryonic object of ~0.6 AU.
26. The timing of the epoch of t ~ $10^{-6}$ s appears confirmed by the WMAP observations of galaxy clustering, because then the subatomic properties survived also and this would be the time of re-constitution into protons, etc.
27. The timing of our t =0 epoch is more exactly the one of the above short flash of photons observed by WMAP, which appears confirmed by WMAP observations of a radiation signature with a wider curvature than that of the 3-K radiation. It is explained by a short burst of re-energized photons before re-constitution of protons, neutrons, and electrons as scatterers to bring high opacity.
28. The present modeling is confirmed by the fact that the dark energy appears to have initiated the expansion of intergalactic space.
29. Because the universes begin at ~ $10^{-6}$ secs instead of t = 0, all problems of the earliest standard theories for Big Bang and atomic theory are gone.
30. New disciplines appear in opportunities for pursuing problems, modeling, predictions, and suggestions for future work.

**Acknowledgement**
I thank Kees de Jager for pointing me towards the Schwarzschild limit.

Alfvén H. & Arrhenius, G.:1976. Evolution of the Solar System, NASA SP-345, Washington D.C.
Basu, D., 2006. QSO pairs across active galaxies: Evidence of blueshifts?, J. Astrophys. & Astron. 27, 381-388, and references there-in.
Beekman, G., 2006, I.O.Yarkovsky and the Discovery of 'his' Effect, J. Hist. Astron. 37, 71-86.
Berkeland, D. J., Hinds, E. A. et al., 1995. Precise Optical Measurement of Lamb Shifts in Atomic Hydrogen. Phys. Rev. Letters 75(13), pp. 2470-2473.
Bitbol, M. (Ed.), 1995. The Schrödinger interpretation of Quantum Mechanics. OxBow Press, Woodbridge, CT.
Bonnell, I. A., Vine, S. G. et al., 2004. Massive star formation: Nurture, not nature. Mon. Not. Roy. Astron. Soc. *349*(2), pp. 735-741.

Bromm, V., 2006. Out of the Dark Ages: the First Stars. Sky & Tel. 111(5), pp. 30-35.
Carr, B. J. & Rees, M. J., 1979. The anthropic principle and the structure of the physical world. Nature 278, pp. 605-612.
Chandrasekhar, S., 1937. The Cosmological Constants. Nature 139, pp. 757-758.
Chandrasekhar S., 1951. The structure, the composition, and the source of energy of the stars. In J. A. Hynek (Ed.), Astrophysics, a Topical Symposium, McGraw-Hill, New York, pp. 508-681.
Chandrasekhar, S., 1987. Truth and Beauty: Aesthetics and Motivations in Science, Univ. Chicago Press, Chicago, IL.
Chandrasekhar, S., 1989. The Cosmological Constants, Selected Papers, Vol. I. Stellar Structure and Stellar Atmospheres, Univ. Chicago Press, Chicago, IL, p. 304.
Copi, C. J., Schramm D. N. et al., 1995. Big-Bang nucleosynthesis and the Baryon Density of the Universe, Science 267, pp. 192 –199.
Cox, A. N. (Ed.), 2000. Allen's Astrophysical Quantities, 4$^{th}$ ed., Springer Verlag, New York, NY, 579.
Dodd, S. 2008. Making Space for Time. Sc. American 298, No. 1, 26-29.
Everett III, H., 1971. The theory of the universal wave function. In B. S. DeWitt & N. Graham, (Eds.) The Many-Worlds Interpretation of Quantum Mechanics, Princeton Univ. Press, Princeton, NJ, pp. 3-140.
Gehrels, T., 2007a. On the Glassy Sea, in Search of a Worldview, Amazon-BookSurge, Charleston, S.C., originally published by Am. Inst. Phys., New York, NY (1988).
Gehrels, T., 2007b. Universes seen by a Chandrasekhar equation in stellar physics. <http://arXiv.org/astro-ph/0701344>.
Gehrels, T., 2007c. The Multiverse and the Origin of our Universe, <http://arXiv.org/abs/0707.1030>.
Gehrels, T., 2007d. Survival through Evolution, from Multiverse to Modern Society, Amazon-BookSurge, Charleston, S.C.
Karshenboim, S. G., 2000. The hydrogen Lamb shift and the proton radius. <http://arXiv:hep-ph/0008137>.
Kleczek, J., 1976. The universe. Reidel Publ., Dordrecht. Netherlands.
Lamb, W. E., Jr., 1995. Anti-photon, Appl. Phys. B 60, pp. 77-84.
Lineweaver, C. H. & Davis, T. M., 2005. Misconceptions about the Big Bang, Sc. American 292(3), pp. 36-45.
Mohr, P.. J. & Taylor, B. N., 2005. CODATA recommended values of the fundamental physical constants: 2002. Rev. Mod. Phys., 80, pp. 633-730, Table XXVI.
Ouchi, M, Hamana, T. et al., 2005. Definitive identification of the transition between small- and large-scale clustering for Lyman break galaxies, Astrophys. J. 635, pp. L117-L120.
Perlmutter, S. et al. 1999. Measurements of Omega and Lambda from 42 high redshift supernovae. Astrophys. J. 517, 565-586.
Planck, M., 1899. Über irreversible Strahlungsvorgänge. Sitzungsber. preusz. Akad. Wissenlschaften 5, pp. 440-490.
Riazuelo, A., Uzan, J.-P. et al., 2004. Simulating cosmic microwave background maps in multiconnected spaces. Phys. Rev. D 69(10), pp. 103514-103542.
Riess, A.G. et al., 1998. Observational evidence from supernovae for an accelerating universe and a cosmological constant. Astron. J. 116. 1009-1038.
Sakharov, A.D., 1965. Zh. Eksp. Teor. Fiz. 49, p. 345; The Initial Stage of an Expanding Universe and the Appearance of a Nonuniform Distribution of Matter. J. Exper. Theor. Phys. 22, p. 241 (1966).
Schweizer, F., 2000. Interactions as a driver of galaxy evolution. Phil. Trans. Roy. Soc. A 358(1772), pp. 2063-2076.
Spergel, D. N., Bean, R., et al., 2007, Wilkinson Microwave Anisotropy Probe (WMAP) Three-Year Results: Implications for Cosmology. Astrophys. J. Suppl. 170(2), pp. 377-470; updated with Five-Year Results in G. Hinshaw, J. L. Wetland, R. S. Hill et al., 2008. <http://wmap.gsfc.nasa.gov>. Stahler, S. W., Palla, F. et al., 2000. The formation of massive stars. In V. Mannings, A. P. Boss & S. S. Russell (Eds.), Protostars and Planets IV, Univ. Ariz. Press, Tucson, AZ, pp. 327-351.

Vilenkin, A., 2006. Many Worlds in One, the Search for Other Universes, Hill and Wang, New York, NY.

**Author Biography**

**Tom Gehrels** is a professor at the University of Arizona where he teaches and does most of his research while he is also a Fellow at the Physical Research Laboratory in Ahmedabad, India, where he teaches in a UN Course for graduates from a variety of Asian countries. He has a Leiden B.S. (Astronomy and Physics, 1951) and Chicago Ph.D. (Astronomy and Astrophysics, 1956). His research has been on asteroids, planets, stars, interstellar medium, and it is presently on universal evolution.